\documentclass{article}
\usepackage{locata,amsmath,graphicx,url,times}
\usepackage{amsfonts,comment}

\newcommand{\norm}[1]{\left\lVert#1\right\rVert}

\pdfoutput=1

\title{Description of algorithms for Ben-Gurion University Submission to the LOCATA challenge}

\name{Lior Madmoni, Hanan Beit-On, Hai Morgenstern and Boaz Rafaely}
\address{Department of Electrical and Computer Engineering\\Ben-Gurion University of the Negev\\Beer-Sheva 84105, Israel\\liomad@gmail.com; hananbo26@gmail.com; haimorg@post.bgu.ac.il; br@bgu.ac.il}
\begin{document}

\ninept
\maketitle

\begin{sloppy}

\begin{abstract}
	This paper summarizes the methods used to localize the sources recorded for the LOCalization And TrAcking (LOCATA) challenge. The tasks of stationary sources and arrays were considered, i.e., tasks 1 and 2 of the challenge, which were recorded with the Nao robot array, and the Eigenmike array. For both arrays, direction of arrival (DOA) estimation has been performed with measurements in the short time Fourier transform domain, and with direct-path dominance (DPD) based tests, which aim to identify time-frequency (TF) bins dominated by the direct sound. For the recordings with Nao, a DPD test which is applied directly to the microphone signals was used. For the Eigenmike recordings, a DPD based test designed for plane-wave density measurements in the spherical harmonics domain was used. After acquiring DOA estimates with TF bins that passed the DPD tests, a stage of k-means clustering is performed, to assign a final DOA estimate for each speaker.
\end{abstract}


%MD-DPD
%\input{MD_DPD}
\section{DOA estimation with the Nao robot array}
\label{sec:NAO_description}
This section describes the method for direction of arrival (DOA) estimation in tasks 1 and 2, that was performed with the Nao robot array.

In this paper, the same spherical coordinate system is used, as described in \cite{lollmann2018locata}, denoted by $(r,\theta,\phi)$, where $r$ is the distance from the origin, and $\theta$ and $\phi$ are the elevation and azimuth angles, respectively. Consider an array of $Q$ omni-directional microphones, representing the array mounted on Nao. In this case, let $\{\mathbf{r}_q\equiv(r_q,\theta_q,\phi_q)\}_{q=1}^{Q}$ denote the microphones positions arranged according to the configuration used in the LOCATA challenge for Nao \cite{lollmann2018locata}.
In addition, a sound field which is comprised of $L$ far field sources is also considered,
arriving from directions $\{\Psi_{l}\equiv(\theta_{l},\phi_{l})\}_{l=1}^{L}$. These $L$ sources
can represent the direct sound from speakers in a room and the reflections
due to objects and room boundaries. In this case, the sound
pressure measured by the array can be described in the short-time Fourier transform (STFT) domain
as \cite{van2002optimum}
\begin{align}
\mathbf{p}(\tau,\omega)=\mathbf{V}(\omega,\mathbf{\Psi})\mathbf{s}(\tau,\omega)+\mathbf{n}(\tau,\omega),\label{eq:pressureModel_vantrees-1}
\end{align}
where $\mathbf{p}(\tau,\omega)=\begin{bmatrix}
p(\tau,\omega,\mathbf{r}_1),p(\tau,\omega,\mathbf{r}_2),\ldots,p(\tau,\omega,\mathbf{r}_Q)
\end{bmatrix}^T$ is a $Q\times1$ vector holding the recorded sound pressure, $\mathbf{s}(\tau,\omega)=\begin{bmatrix}s_{1}(\tau,\omega),s_{2}(\tau,\omega),\ldots,s_{L}(\tau,\omega)\end{bmatrix}^{T}$
is an $L\times1$ vector holding the source signal amplitudes, $\mathbf{V}(\omega,\mathbf{\Psi})$
is a $Q\times L$ matrix holding the steering vectors between
each source and microphone and with $\mathbf{\Psi}=\begin{bmatrix}
\Psi_{1} , \Psi_{2} ,\ldots, \Psi_{L}
\end{bmatrix}^T$ denoting the DOAs of the sources, $\mathbf{n}(\tau,\omega)=\begin{bmatrix}n_{1}(\tau,\omega),n_{2}(\tau,\omega),\ldots,n_{Q}(\tau,\omega)\end{bmatrix}^{T}$
is a $Q\times1$ vector holding the noise components, $\tau$ and
$\omega$ are the time and frequency indices, respectively, and
$(\cdot)^{T}$ denotes the transpose operator.

\begin{comment}

The microphone domain direct-path dominance test (MD-DPD) first identifies
time-frequency (TF) bins in the STFT domain for which the direct-path
from a single speaker is dominant. Then, MUSIC \cite{MUSIC} with a signal subspace
of single dimension is applied to each of the selected bins yielding
DOA estimation for each bin. Finally, the DOA estimations from the
different bins are fused together using K-means to obtain the final
DOAs estimation. 
\end{comment}

The signals recorded by the Nao robot array were transformed
to the STFT domain with a Hanning window of 512 samples (32 ms), and
with an overlap of 50\%. A focusing process was then applied to this
measured pressures vector in order to remove the frequency dependence
of the steering matrices across every $J_{\omega}=15$ adjacent frequency
indexes. The purpose of the focusing process is to enable the implementation
of frequency-smoothing while preserving the spatial information. The
focusing was performed by multiplying the sound pressure vector at each
frequency index, $\omega$, with a focusing transformation $\mathbf{T}\left(\omega,\omega_{0}\right)$
that satisfies

\begin{equation}
%\mathbf{T}\left(\omega,\omega_{0}\right)\mathbf{V}\left(\mathbf{\psi},\omega\right)=\mathbf{V}\left(\mathbf{\psi},\omega_{0}\right),\label{eq:focusing eq-1}
\mathbf{T}\left(\omega,\omega_{0}\right)\mathbf{V}(\omega,\mathbf{\Psi})=\mathbf{V}(\omega_0,\mathbf{\Psi}),\label{eq:focusing eq-1}
\end{equation}
where $\omega_{0}$ is the center frequency in the frequency-smoothing
range. The focusing transformations were computed in advance according
to \cite{MD_DPD_Submitted} using spherical harmonics (SH) order of $N=4$. With ideal
focusing, the cross-spectrum matrix of the focused sound pressure can be
written as \cite{MD_DPD_Submitted}

\begin{equation}
\mathbf{S}_{\tilde{p}}\left(\tau,\omega\right)=\mathbf{V}(\omega_0,\mathbf{\Psi})\mathbf{S}_{s}\left(\tau,\omega\right)\mathbf{V}(\omega_0,\mathbf{\Psi})^{H}+\mathbf{S}_{\widetilde{n}}\left(\tau,\omega\right)\label{eq:focused pressure cross-spectrum}
\end{equation}
where $\mathbf{S}_{\tilde{p}}\left(\tau,\omega\right)=E\left[\mathbf{T}\left(\omega,\omega_{0}\right)\mathbf{p}\left(\tau,\omega\right)\mathbf{p}\left(\tau,\omega\right)^{H}\mathbf{T}\left(\omega,\omega_{0}\right)^{H}\right]$,
$\mathbf{S}_{s}\left(\tau,\omega\right)=E\left[\mathbf{s}\left(\tau,\omega\right)\mathbf{s}\left(\tau,\omega\right)^{H}\right]$,
$\mathbf{S}_{\tilde{n}}\left(\tau,\omega\right)=E\left[\mathbf{T}\left(\omega,\omega_{0}\right)\mathbf{n}\left(\tau,\omega\right)\mathbf{n}\left(\tau,\omega\right)^{H}\mathbf{T}\left(\omega,\omega_{0}\right)^{H}\right]$, and $(\cdot)^{H}$ is the Hermitian operator.
In practice, an averaging across $J_{\tau}=3$ time frames is used
to approximate the expectation. A frequency-smoothing is then applied
to $\mathbf{S}_{\tilde{p}}\left(\tau,\omega\right)$ by averaging
across $J_{\omega}=15$ frequency bins. Denoting the smoothed variables
by an overline, i.e. $\overline{\mathbf{S}}\left(\tau,\omega\right)=\sum_{j_{\omega}=0}^{J_{\omega}-1}\mathbf{S}\left(\tau,\omega-j_{\omega}\right)$,
the smoothed focused cross-spectrum matrix can be written as

\begin{equation}
\overline{\mathbf{S}_{\tilde{p}}}\left(\tau,\omega\right)=\mathbf{V}\left(\mathbf{\psi},\omega_{0}\right)\overline{\mathbf{S}_{s}}\left(\tau,\omega\right)\mathbf{V}\left(\mathbf{\psi},\omega_{0}\right)^{H}+\overline{\mathbf{S}_{\widetilde{n}}}\left(\tau,\omega\right).\label{eq:smoothed focused cross-spectrum}
\end{equation}
The purpose of the frequency-smoothing operation is to restore the
rank of the source cross-spectrum matrix, $\mathbf{S}_{s}\left(\tau,\omega\right)$,
which is singular when coherent sources, such as reflections, are
present. After applying focusing and frequency-smoothing, the effective-rank
\cite{roy2007effective} of $\overline{\mathbf{S}_{\tilde{p}}}\left(\tau,\omega\right)$
reflects the number of sources $Q$ and the noise subspace can be
correctly estimated \cite{MD_DPD_Submitted}. Time-frequency (TF) bins in which the direct-path
is dominant are identified in a similar way to those proposed in the direct-path dominance (DPD)
test \cite{nadiri2014localization} 

\[
\mathfrak{\mathcal{A}}_{\text{MD-DPD}}=\left\{ \left(\tau,\nu\right):\frac{\lambda_{1}\left(\overline{\mathbf{S}_{\tilde{p}}}\left(\tau,\omega\right)\right)}{\lambda_{2}\left(\overline{\mathbf{S}_{\tilde{p}}}\left(\tau,\omega\right)\right)}>\mathcal{TH}_{\text{MD-DPD}}\right\} ,
\]
where $\lambda_{1}\left(\overline{\mathbf{S}_{\tilde{p}}}\left(\tau,\omega\right)\right)$
and $\lambda_{2}\left(\overline{\mathbf{S}_{\tilde{p}}}\left(\tau,\omega\right)\right)$
are the largest and the second largest eigenvalues of $\overline{\mathbf{S}_{\tilde{p}}}\left(\tau,\omega\right)$,
and $\mathcal{TH}_{\text{MD-DPD}}$ is the test threshold, chosen independently
for each recording, to ensure that 5\% of all available bins pass
the test. Then, MUSIC with a signal subspace of single dimension was
applied to each of the bins in $\mathcal{A}_{\text{MD-DPD}}$. The noise
subspace was estimated by the singular values decomposition of $\overline{\mathbf{S}_{\tilde{p}}}\left(\tau,\omega\right)$.

Next, k-means clustering was performed with the DOA estimates from
the bins that passed the test. For task 1, a single speaker was present,
thus, k-means clustering has been performed with a single cluster.
For task 2, the number of clusters was chosen to the number of sources,
which has been estimated for each recording by examining the scatter
of DOA estimates on an azimuth-elevation grid, and was therefore assumed
to be known apriori. This was performed in order to focus on the performance
of the DOA estimation process rather than on source number estimation. Finally, since the sources in tasks
1 and 2 are known to be stationary, the final DOA estimates have been
associated with a unique source identifier for all timestamps, regardless
of its activity.

%DPD-EDS
%\input{DPD_EDS}
\section{DOA estimation with the Eigenmike array}
\label{sec:EM_description}
This section describes the method for DOA estimation in tasks 1 and 2, that was performed with the Eigemike array.

%System model - anm
The sound pressure system model described in (\ref{eq:pressureModel_vantrees-1}), can be used with $r_q=r$ for all $q=1,\ldots,Q$ and with the same STFT parameters, such that it now describes a spherical array. This formulation can facilitate the processing of signals in the SH domain \cite{SHdomain_cite1,SHdomain_cite2,rafaely2015fundamentals}, which was performed up to SH order of $N=3$. Following that, plane wave decomposition had been performed, leading to \cite{khaykin2009coherent}:
\begin{align}
\mathbf{a_{nm}}(\tau,\omega)=\mathbf{Y}^H(\mathbf{\Psi}) \mathbf{s}(\tau,\omega)+\tilde{\mathbf{n}}(\tau,\omega),
\label{eq:SHPWDModel}
\end{align}
where $$\mathbf{a_{nm}}(\tau,\omega)=\begin{bmatrix}
a_{00}(\tau,\omega),a_{1(-1)}(\tau,\omega),a_{10}(\tau,\omega),\ldots,a_{NN}(\tau,\omega)
\end{bmatrix}^T$$ is a $(N+1)^2\times 1$ vector holding the recorded plane wave density (PWD) coefficients in the SH domain, $
\mathbf{Y}^H(\mathbf{\Psi})=\begin{bmatrix}
\mathbf{y}^*(\Psi_1) , \mathbf{y}^*(\Psi_2) , \ldots , \mathbf{y}^*(\Psi_L)
\end{bmatrix}%\in\mathbb{C}^{Q\times (N+1)^2},
$ is the $(N+1)^2\times L$ steering matrix in this domain, with its columns
$
\mathbf{y}(\Psi_l)=\big[Y_0^0(\Psi_l),Y_{1}^{-1}(\Psi_l),\ldots, Y_{N}^{N}(\Psi_l)\big]^T,%\in\mathbb{C}^{(N+1)^2\times 1},
$
holding the SH functions $Y_n^m(\cdot)$ of order $n$ and degree $m$. These functions are assumed to be order limited to $N$, which usually holds when both $N=\lceil kr\rceil$ and $(N+1)^2\leq Q$ \cite{truncation_n_ceil_kr,rafaely2015fundamentals}, where $k$ is the wavenumber. The noise components in this domain are described by the $(N+1)^2\times 1$ vector $\tilde{\mathbf{n}}(\tau,\omega)$, where $(\cdot)^*$ denotes the complex conjugate. In this challenge, this plane-wave decomposition was performed in a similar manner to the R-PWD method, described in \cite{Robust_PWD_AlonRafa} (equation (2.27)).

%System model - correlation matrices
Next, the local TF correlation matrices are computed for every TF bin by \cite{nadiri2014localization}: 
%averaging the PWD measurements over $J_{\tau}$ time frames and $J_{\omega}$ frequency bins:
\begin{align}
\tilde{\mathbf{S}}_a(\tau,\omega)=&\frac{1}{J_{\tau}J_{\omega}}\sum_{j_{\omega}=0}^{J_{\omega}-1}\sum_{j_{\tau}=0}^{J_{\tau}-1}\mathbf{a_{nm}}(\tau-j_{\tau},\omega-j_{\omega})\nonumber\\
&\times\mathbf{a_{nm}}^H(\tau-j_{\tau},\omega-j_{\omega}),
%& \approx \mathbf{Y}^H(\mathbf{\Psi}){\mathbf{R}}_{\mathbf{s}}(\tau,\omega)\mathbf{Y}(\mathbf{\Psi}) + {\mathbf{R}}_{\tilde{\mathbf{n}}}(\omega),
\label{eq:Ra_Model}
\end{align}
where $J_{\tau}$ and $J_{\omega}$ are the number of time and frequency bins for the averaging, respectively. The values that were chosen for this array are  $J_{\tau}=2$ and $J_{\omega}=15$.
Notice in (\ref{eq:Ra_Model}) that frequency smoothing is performed directly without focusing matrices, in this domain \cite{FSdecor2_SH}.

%DPD-EDS test
The direct-path dominance enhanced plane-wave decomposition (DPD-EDS) test is designed for PWD measurements in the SH domain, and it uses the local TF correlation matrix $\tilde{\mathbf{S}}_a(\tau,\omega)$, as in (\ref{eq:Ra_Model}). With the aim of identifying TF bins dominated by the direct sound, it was shown in \cite{DPD_EDS_Submitted}, that under some conditions, the dominant eigenvector of $\tilde{\mathbf{S}}_a(\tau,\omega)$, denoted by $\mathbf{u}_1(\tau,\omega)$, may approximately satisfy
\begin{align}
\mathbf{u}_1(\tau,\omega) \propto \mathbf{y}^*(\Psi_1),
\label{eq:u1}
\end{align}
where $\Psi_1$ is the direction of the direct sound in the TF bin. % These condition are met when the direct sound and the reflections are sufficiently spatially separated, and when the sources are uncorrelated. This will approximately hold for a direct sound and its reflection after frequency smoothing, which has been shown to perform de-correlation \cite{FSdecor1,FSdecor2_SH}, and which is performed in (\ref{eq:Ra_Model}).  
Motivated by (\ref{eq:u1}), identifying a bin dominated by the direct sound, can be achieved by examining $\mathbf{u}_1(\tau,\omega)$, and measuring to what extent it represents a single plane wave. In this challenge, this has been performed by the following MUSIC-based measure
\begin{align}
\mathcal{EDS}(\tau,\omega)=\underset{\Omega}{\text{max}}\,\frac{1}{\norm{   \mathbf{P}_{\mathbf{u}_{1}(\tau,\omega)}^{\perp}\mathbf{y}^{*}(\Omega)  }^2},
\label{eq:EDS_measure_MUSIC}
\end{align}
where $\mathbf{P}_{\mathbf{u}_{1}(\tau,\omega)}^{\perp}$ 
%\begin{align}
%\mathbf{P}_{\mathbf{u}_{1}(\tau,\omega)}^{\perp} = \mathbf{I} - \frac{ \mathbf{u}_{1}(\tau,\omega) \mathbf{u}_{1}^H(\tau,\omega) }{ \norm{ \mathbf{u}_{1}(\tau,\omega) }^2 }
%\end{align}
is the projection into the subspace which is orthogonal to $\mathbf{u}_{1}(\tau,\omega)$.
Next, the following DPD-EDS test have been performed:
\begin{align}
\mathcal{A}_{\text{EDS}}=\Big\{  (\tau,\omega): \mathcal{EDS}(\tau,\omega)>\mathcal{TH}_{\text{EDS}}  \Big\},
\label{eq:EDS_thr}
\end{align}
where $\mathcal{TH}_{\text{EDS}}$ is the test thresholds which should hold $\mathcal{TH}_{\text{EDS}}\gg 1$, and in this challenge was chosen for each recording separately, to ensure that $2.5\%$ of all available bins pass the test. %This is because the measure suggested in (\ref{eq:EDS_measure_MUSIC}) returns zero in the denominator for perfect similarity to a single plane wave, and so, an expected high score for the measure should satisfy $\mathcal{TH}_{\text{EDS}}\gg1$. 

%DOA est - bin-wise
Similarly to the previous section, a DOA estimation from each TF bin is given by the argument $\Omega$ that maximizes $\mathcal{EDS}(\tau,\omega)$, 
\begin{align}
\Omega_{\text{EDS}}=\big\{ \Omega :\text{arg}\, \underset{\Omega}{\text{max}}\,\mathcal{EDS}(\tau,\omega), \forall (\tau,\omega)\in\mathcal{A}_{\text{EDS}} \big\},
\label{eq:DOAest_EDS}
\end{align}
already computed in (\ref{eq:EDS_measure_MUSIC}). For further information on the DPD-EDS test, the reader is referred to \cite{DPD_EDS_Submitted,DPD_SPW_EUSIPCO}.
The process of producing the final DOA estimates is performed similarly to the process described for the Nao robot array in the previous section, using k-means clustering.

For most recordings, an analysis frequency range of $[400,6000]\,$Hz was employed, with the exception of several recordings where the frequency range was reduced to $[400,4000]\,$Hz which seemed to yield more tightly dense clusters of DOA estimates. 
%bias
When the development data of the Eigenimke recordings was analyzed, a relatively constant bias of $+8^{\circ}$ in the azimuth angle, and $-5^{\circ}$ in the elevation angle, relative to the ground truth data, was present. Hence, this bias was subtracted from the final DOA estimates that were calculated with the evaluation data, for all recordings.

%For task 1, the final DOA estimate was then computed as the mean of $\Omega_{\text{EDS}}$. For task 2, the final DOA estimate for each speaker was calculated with a further step of k-means clustering. The number of clusters was chosen as the number of sources, which was assumed to be known a-priori, and the mean of each cluster was calculated as the final DOA estimate for each speaker.

% -------------------------------------------------------------------------
% Either list references using the bibliography style file IEEEtran.bst
\bibliographystyle{IEEEtran}
\bibliography{LOCATA_refs}

\end{sloppy}
\end{document}